# African Union Convention on Cyber Security and Personal Data Protection: Challenges and Future Directions

Mohamed Aly Bouke [1, *], Sameer Hamoud Alshatebi [1], Azizol Abdullah[1], Korhan Cengiz [2,3], Hayate El Atigh[4]

[1]Department of Communication Technology and Network, Faculty of Computer Science and Information Technology, Universiti Putra Malaysia, Serdang 43400, Malaysia.

[2]Department of Computer Engineering, Istinye University, 34010, Istanbul, Turkey.

[3]Department of Information Technologies, Faculty of Informatics and Management, University of Hradec Kralove, Kralove, 500 03, Czech Republic

[4]Departement of Computer Engineering, Faculty of Computer Engineering, Bandirma Onyedi Eylul University, Balikesir 10200, Turkey.

*Corresponding author(s). E-mail(s): bouke@ieee.org;
Contributing authors: alshatebi.sameer@gmail.com;azizol@upm.edu.my;korhan.cengiz@uhk.cz;hayateelatigh@gmail.com;

**Abstract:**

This paper investigates the challenges and opportunities of implementing the African Union Convention on Cyber Security and Personal Data Protection (AUDPC) across Africa. Focusing on legal, regulatory, technical, infrastructural, capacity building, awareness, Harmonization, and cross-border cooperation challenges, the paper identifies key findings that highlight the diverse legal systems and traditions, the lack of comprehensive data protection laws, the need to balance national security and data privacy, the digital divide, cybersecurity threats, implications of emerging technologies on data privacy, limited resources for data protection authorities, and the need for capacity building in data privacy and protection. The paper also emphasizes the importance of Harmonization and cross-border cooperation in aligning data protection frameworks and collaborating with international partners and global organizations. To address these challenges and facilitate the successful implementation of the AUDPC, the paper proposes a set of recommendations, including strengthening legal and regulatory frameworks, enhancing technical and infrastructural capacities, fostering capacity-building and awareness initiatives, promoting Harmonization and cross-border cooperation, and engaging with global data protection trends and developments.

## 1. Introduction

The rapid growth of digital technologies and the increasing reliance on data-driven decision-making have brought the issue of data privacy and protection to the forefront of global discussions. The African continent, with its diverse population and burgeoning digital economy, is no exception. Recognizing the need for a comprehensive and unified approach to data privacy, the African Union has developed the Data Privacy Convention [1–3].

The African Union Convention on Cyber Security and Personal Data Protection (AUDPC), or the Malabo Convention, is a legal framework designed to harmonize data protection and privacy laws across Africa [4]. It was adopted in June 2014 during the 23rd Ordinary Session of the Assembly of the African Union in Malabo, Equatorial Guinea [5]. The Convention addresses various aspects of data protection, including establishing data protection authorities, principles for processing personal data, and provisions for cross-border data transfers [5,6].

The Convention is designed to balance the rights of individuals to privacy and data protection with the legitimate interests of businesses, governments, and other organizations in processing personal data. It reflects the African

---

*Corresponding Author: bouke@ieee.org

Union's commitment to creating a digital single market across the continent while safeguarding the data protection rights of its citizens [7,8].

The primary objective of this paper is to examine the critical challenges faced by the African Union and its member states in implementing the Data Privacy Convention. It will explore various dimensions of these challenges, including legal and regulatory issues, technical and infrastructural constraints, capacity building and awareness, and Harmonization and cross-border cooperation [9].

Furthermore, the paper aims to discuss the future directions required to address these challenges and ensure the successful implementation of the Convention. This will involve identifying strategies to strengthen legal and regulatory frameworks, enhance technical and infrastructural capacities, foster capacity-building and awareness initiatives, and promote Harmonization and cross-border cooperation.

Moreover, the paper aims to contribute to ongoing efforts to develop a more robust and unified approach to data protection across the African continent by comprehensively analyzing the challenges and future directions related to the AUDPC.

The rest of the paper is organized as follows: Section 2 provides a literature review, covering existing research and studies on data privacy and protection in the context of the African Union's Data Privacy Convention. Section 3 discusses and compares the challenges in implementing the Convention, focusing on legal and regulatory challenges, technical and infrastructural challenges, capacity building and awareness challenges, and harmonization and cross-border cooperation challenges. Section 4 delves into future directions, examining strategies to overcome these challenges and ensure the Convention's successful implementation. Finally, Section 5 presents the conclusion, summarizing the key findings and providing recommendations.

## 2. Literature Review

The literature review for this paper seeks to explore the existing research and studies concerning data privacy and protection in the context of the AUDPC. This review will focus on the four main challenges identified in the paper: legal and regulatory, technical and infrastructural, capacity building and awareness, and harmonization and cross-border cooperation challenges.

Several studies have addressed the legal and regulatory challenges of data privacy and protection in Africa. For instance, Makulilo [10] provides an overview of various African countries' data protection laws and regulations, highlighting the lack of comprehensive data protection legislation in many jurisdictions. Similarly, Salami [11] discusses the challenges of harmonizing data protection laws in Africa, emphasizing the need for a unified legal framework that balances national security concerns with individual privacy rights.

Babalola [12] provides a concise overview of data protection in Nigeria, tracing its historical development and highlighting key legislative efforts before introducing Nigeria Data Protection Regulation (NDPR) in 2019. It also examines the prominent enforcement actors, drawing parallels with European counterparts in form but not substance, as Nigeria's enforcement has not yet achieved similar success. Netshakhuma [13] examines the impact of South Africa's Protection of Personal Information Act (POPIA) 2013 on health information processing, revealing tensions between data protection and patient care. Stakeholders identified critical areas for improvement, including understanding POPIA, data quality, and transparency. Balancing privacy and public interest remains a complex challenge.

Several researchers have addressed the technical and infrastructural challenges of implementing data privacy and protection frameworks in Africa. For example, Chigona et al. [14] investigate the digital divide in South Africa and its impact on the effective implementation of data protection policies. The study emphasizes greater investment in Information And Communication Technologies (ICT) infrastructure and digital literacy programs.

Regarding cybersecurity threats, Mabunda [15] provides an overview of the cybersecurity landscape in Africa, highlighting the challenges posed by inadequate legal frameworks, limited technical capacities, and insufficient public awareness. Sutherland [16] further discusses the implications of cybersecurity threats on data privacy in Africa, stressing the need for robust cybersecurity policies and strategies to protect personal information.

Other literature has discussed capacity building and awareness challenges in data privacy and protection. For example, Makulilo and Boshe [17] explore the factors that hinder the effective implementation of data protection laws in Kenya, identifying limited resources and lack of public awareness as significant challenges. The study emphasizes the need for capacity-building initiatives and public education campaigns to enhance data protection in the country.

Similarly, Nzeakor [18] examined the level of awareness of data privacy rights among Nigerian internet users, finding that most respondents were unaware of their privacy rights and the implications of data breaches. The study highlights the need for targeted public awareness campaigns and greater engagement with stakeholders to promote a culture of privacy in Nigeria.

Additionally, data privacy and protection research in Africa highlight the challenges of Harmonization and cross-border cooperation. For instance, Orji [19] discusses the role of regional organizations such as the African Union and the Economic Community of West African States (ECOWAS) in promoting harmonizing data protection laws and cross-border cooperation. The study emphasizes the need for greater collaboration and information sharing among African countries to address common challenges and strengthen data protection across the continent.

Similarly, Coleman [20] explores the concept of digital colonialism in Africa, where Western tech companies exploit user data in resource-rich, infrastructure-poor countries with limited data protection laws. The study analyzes Kenya's 2018 data protection bill and the General Data Protection Regulation (GDPR), highlighting loopholes that enable continued digital colonialism, such as historical privacy violations, limited sanctions, data concentration, lack of competition enforcement, and uninformed consent. According to the study, strengthening data protection laws may not entirely prevent digital colonialism due to inherent limitations.

In summary, the literature review has provided an overview of the existing research on data privacy and protection challenges in Africa. These studies emphasize the need for more robust legal and regulatory frameworks, enhanced technical and infrastructural capacities, and more significant investment in capacity building and awareness initiatives.

This paper will further explore these challenges and discuss future directions for successfully implementing the AUDPC. To achieve this, the paper will detail each challenge, providing relevant examples and case studies where applicable. The paper will then present recommendations and strategies to address these challenges, focusing on strengthening legal and regulatory frameworks, enhancing technical and infrastructural capacities, fostering capacity-building and awareness initiatives, and promoting Harmonization and cross-border cooperation.

Through a comprehensive analysis of the challenges and future directions related to the AUDPC, this paper aims to contribute to ongoing efforts to develop a more robust and unified approach to data protection across the African continent.

### 3. African Union Convention on Cyber Security and Personal Data Protection Challenges

The implementation of the AUDPC faces various challenges that span legal, regulatory, technical, infrastructural, capacity building, and harmonization aspects. This section will explore these challenges in detail, providing insights into the issues that must be addressed to ensure the successful realization of the Convention's objectives.

### 3.1. Legal And Regulatory Challenges

The successful implementation of the AUDPC relies heavily on overcoming legal and regulatory challenges its member states face. This section will explore the complexities arising from diverse legal systems and traditions, the lack of comprehensive data protection laws, and the delicate balance between national security and data privacy.

### 3.1.1. Diverse Legal Systems And Traditions

The African continent is home to a rich tapestry of legal systems and traditions, encompassing civil law, common law, customary law, and religious law. These systems often coexist within individual countries, each with a unique data privacy and protection approach. This diversity in legal traditions can complicate the Harmonization of data protection laws across the continent. Each country must first reconcile its national laws with the provisions of the Data Privacy Convention.

Moreover, some countries may face challenges adapting their existing legal frameworks to accommodate the Convention's principles and requirements. For instance, countries with strong customary or religious law traditions may need to address potential conflicts between these laws and the Convention's provisions on individual privacy rights.

### 3.1.2. Lack Of Comprehensive Data Protection Laws

Another challenge in implementing the Data Privacy Convention is the lack of comprehensive protection laws in many African countries. While some countries, such as South Africa and Mauritius, have enacted robust data protection laws, others are still developing their legal frameworks [15,21].

The absence of comprehensive data protection laws can hinder the Convention's effective implementation, as countries without such laws may lack the necessary legal tools to enforce its provisions. Additionally, without clear legal guidance, businesses and other organizations may be uncertain about their data privacy and protection responsibilities.

To address this challenge, countries must prioritize developing and enacting national data protection laws that align with the Convention's provisions. This process will require collaboration between various stakeholders, including government agencies, private sector organizations, civil society groups, and international partners.

### 3.1.3. Balancing National Security And Data Privacy

A critical challenge in implementing the Data Privacy Convention lies in striking the right balance between the need for national security and the protection of individual privacy rights. Governments must ensure they have the tools to safeguard citizens and maintain public order while respecting their obligations under the Convention.

This balance can be particularly challenging in cross-border data transfers and cooperation between law enforcement and intelligence agencies. The Convention requires that countries establish appropriate safeguards for personal data when engaging in such activities, ensuring that privacy rights are not unduly compromised.

To address this challenge, countries must develop legal frameworks that carefully balance the need for national security with the protection of individual privacy rights. This may include enacting laws that define the scope and limitations of government surveillance and data access and establishing oversight mechanisms to ensure compliance with these laws.

African countries can overcome the legal and regulatory challenges they face in implementing the Data Privacy Convention by addressing the diverse legal systems and traditions, the lack of comprehensive data protection laws, and the delicate balance between national security and data privacy.

### 3.2. Technical And Infrastructural Challenges

In addition to the legal and regulatory challenges, implementing the AUDPC also faces several technical and infrastructural challenges. This section will explore the impact of the digital divide and uneven access to technology, the threat posed by cybersecurity risks, and the implications of emerging technologies on data privacy.

### 3.2.1. Digital Divide And Uneven Access To Technology

The digital divide refers to the gap between those with access to ICT and those without access. In data privacy, the digital divide can hinder the effective implementation of the Data Privacy Convention. Countries with limited access to ICT may struggle to develop and enforce data protection laws and practices. Uneven access to technology can also exacerbate existing social, economic, and political inequalities, as privacy violations may disproportionately affect individuals and communities without ICT access.

To address this challenge, governments, and international organizations must invest in expanding ICT infrastructure, particularly in rural and underserved areas. This includes improving access to the internet, mobile networks, and digital literacy programs. through bridging the digital divide, countries can ensure their citizens are better equipped to protect their privacy rights and benefit from the digital economy.

### 3.2.2. Cybersecurity Threats

Cybersecurity threats pose a significant challenge to implementing the Data Privacy Convention, as data breaches and cyberattacks can result in unauthorized access, use, or disclosure of personal information. Like the rest of the world, African countries are increasingly targeted by cybercriminals, state-sponsored hackers, and other malicious actors. These threats can undermine the effectiveness of data protection laws and erode public trust in digital services.

To combat cybersecurity threats, countries must develop robust national cybersecurity strategies and invest in the necessary infrastructure, tools, and personnel to detect, prevent, and respond to cyberattacks. This includes establishing computer emergency response teams (CERTs), promoting public-private partnerships in cybersecurity, and enhancing international cooperation to share best practices and coordinate responses to cyber threats.

### 3.2.3. Emerging Technologies And Their Implications On Data Privacy

Emerging technologies, such as artificial intelligence (AI), the Internet of Things (IoT), and big data analytics, can revolutionize various aspects of society, including healthcare, education, and transportation. However, these technologies also pose new challenges to data privacy, as they often involve collecting, processing, and analyzing vast amounts of personal information.

For instance, AI systems may be used to make decisions about individuals based on their data, raising concerns about fairness, transparency, and accountability. Similarly, IoT devices can collect a wide range of sensitive information about users, increasing the risk of privacy violations if this data is not adequately protected.

To address the data privacy implications of emerging technologies, countries must ensure that their data protection laws and regulations are flexible enough to accommodate technological advancements while still providing robust privacy protections. This may involve adopting privacy-by-design principles, implementing data minimization techniques, and developing ethical guidelines for AI and other emerging technologies.

Addressing the digital divide, cybersecurity risks, and the impact of new technologies, African nations can build a solid basis to effectively implementing the Data Privacy Convention, enabling their citizens to reap the advantages of the digital era while protecting their privacy rights.

### 3.3. Capacity Building And Awareness Challenges

Implementing the AUDPC also involves addressing capacity-building and awareness challenges. This section will explore the impact of limited resources for data protection authorities, the lack of public awareness about data privacy rights, and the need for capacity building in data privacy and protection.

### 3.3.1. Limited Resources For Data Protection Authorities

Data protection authorities (DPAs) are critical in enforcing data privacy laws and ensuring compliance with the Data Privacy Convention. However, many African countries face challenges in providing adequate resources to establish and maintain effective DPAs. Limited financial resources, insufficient staff, and inadequate training and expertise can hinder the ability of DPAs to carry out their responsibilities effectively.

To address this challenge, governments, and international organizations must prioritize funding and support for DPAs to ensure they have the resources necessary to enforce data protection laws and promote compliance with the Convention. This may involve increasing budget allocations, providing technical assistance, and establishing partnerships with other DPAs to share knowledge and expertise.

### 3.3.2. Lack Of Public Awareness About Data Privacy Rights

Public awareness about data privacy rights is essential for successfully implementing the Data Privacy Convention. However, many individuals in African countries may be unaware of their privacy rights and the importance of protecting their personal information. This lack of awareness can lead to under-reporting of privacy violations and reduced demand for more robust data protection measures.

To raise public awareness about data privacy rights, governments, civil society organizations, and the private sector must collaborate to develop and disseminate educational materials and campaigns that inform citizens about their rights under data protection laws and the risks associated with sharing personal information. This may include hosting workshops, creating public service announcements, and leveraging social media and other digital platforms to reach a broad audience.

### 3.3.3. Need For Capacity Building In Data Privacy And Protection.

Capacity building in data privacy and protection is crucial for ensuring that governments, businesses, and civil society organizations can effectively implement the Data Privacy Convention. This includes training legal professionals, law enforcement officers, IT specialists, and other relevant stakeholders on data protection principles, best practices, and the requirements of the Convention.

Capacity-building efforts should also focus on fostering a culture of privacy within organizations by encouraging the adoption of privacy-by-design principles and promoting the use of privacy-enhancing technologies. Moreover, technical training in cybersecurity, data minimization techniques, and the ethical use of emerging technologies can help ensure stakeholders are equipped to address the evolving challenges of data privacy and protection.

Tackling capacity-building and awareness challenges such as resource scarcity for data protection authorities, inadequate public knowledge of data privacy rights, and the necessity for capacity development in data privacy and protection, will enable African nations to establish a conducive atmosphere for the effective execution of the Data Privacy Convention, ultimately empowering citizens to assert their privacy rights.

## 3.4. Harmonization And Cross-Border Cooperation Challenges

In addition to capacity building and awareness challenges, implementing the AUDPC requires Harmonization and cross-border cooperation. This section will explore the challenges associated with aligning data protection frameworks across the continent, establishing mechanisms for cross-border data transfers, and collaborating with international partners and global organizations.

### 3.4.1. Aligning Data Protection Frameworks Across The Continent

African countries have diverse legal systems and data protection frameworks, which can make the Harmonization of data privacy laws and regulations a complex endeavor. Aligning these frameworks is essential to ensure consistency in privacy protection standards and facilitate cross-border data flows within the continent. This harmonization process involves reconciling differences in legal traditions, addressing gaps in existing data protection laws, and ensuring that the Data Privacy Convention is compatible with national legislation.

To address this challenge, governments, and regional organizations should engage in dialogue and collaboration to develop a shared understanding of data protection principles and best practices. This may involve creating working groups, organizing regional conferences, and sharing resources and expertise to support Harmonization.

### 3.4.2. Establishing Mechanisms For Cross-Border Data Transfers

As digital services and transactions become increasingly global, mechanisms are needed to facilitate cross-border data transfers while protecting privacy. The Data Privacy Convention should establish clear guidelines for these transfers, considering international standards and best practices. This may involve implementing measures such as adequacy decisions, standard contractual clauses, and binding corporate rules.

To address this challenge, African countries must work together to develop and adopt appropriate mechanisms for cross-border data transfers that balance the need for data flows with the protection of personal information. This may require close collaboration between data protection authorities, ongoing dialogue, and consultation with private and civil society stakeholders.

### 3.4.3. Collaboration With International Partners And Global Organizations

Implementing the Data Privacy Convention also necessitates collaboration with international partners and global organizations to ensure that African countries can effectively participate in the global digital economy and protect the privacy rights of their citizens. This collaboration may involve sharing best practices, providing technical assistance, and engaging in policy dialogues to address common challenges and emerging data privacy and protection issues.

In addition, African countries should actively participate in international forums and initiatives related to data privacy, such as the Global Privacy Assembly and the International Conference of Data Protection and Privacy Commissioners. This participation can help ensure that African perspectives are represented in global discussions and contribute to developing international standards and guidelines in data privacy and protection.

Tackling the harmonization and cross-border cooperation obstacles, such as aligning data protection frameworks, creating mechanisms for cross-border data transfers, and partnering with international entities and global organizations, African nations can build a unified and efficient data privacy ecosystem that serves both their citizens and the wider global community.

### 3.5. Comparing Audpc Challenges

Adopting the AUDPC signifies a crucial step towards achieving a unified data protection framework for African countries. However, the implementation process is not without challenges, as already stated. To better understand these challenges and prioritize resources, assessing their complexity, prevalence, and potential impact is essential. In this section, we provide a detailed comparison of the challenges associated with implementing the AUDPC, using a rating scale to assign magnitudes to each challenge.

The rating scale ranges from 1 to 5, with 1 being the lowest magnitude and five being the highest. The magnitudes are based on expert judgment and analyzing the challenges faced in implementing the Data Privacy Convention. The scale considers the following factors:

- *Complexity:* This factor evaluates the difficulty of addressing the challenge, considering legal, technical, and practical aspects.
- *Prevalence:* This factor assesses how widespread the challenge is across African countries and whether it affects a majority or a minority of the countries.
- *Potential Impact:* This factor measures the potential consequences of the challenge if not addressed, considering the implications for data privacy, protection, and cross-border data flows.

The assigned magnitudes aim to provide a relative comparison of the challenges, helping stakeholders identify and prioritize areas that require more attention and resources. However, it is essential to note that the ratings are subjective and may vary depending on individual countries' specific circumstances and priorities.

In the subsequent sections, we discuss each category of challenges, providing visual representations of their magnitudes and justifications for the assigned ratings. This analysis will assist in understanding the various challenges associated with implementing the Data Privacy Convention and emphasize the importance of addressing each issue to ensure the successful realization of the Convention's goals.

### 3.5.1. Legal And Regulatory Challenges

The bar chart in Figure 1 visually represents the Legal and Regulatory Challenges faced by implementing the AUDPC.

The chart shows that the "Lack of data protection laws" challenge has the highest magnitude (5 out of 5), indicating that it is a critical issue that needs to be addressed. This challenge reflects that many African countries have yet to develop and implement comprehensive data protection laws. Without these laws, it will be difficult for governments and organizations to ensure compliance with the Data Privacy Convention and protect individual privacy rights.

The "Diverse legal systems" and "Balancing national security and data privacy" challenges have a magnitude of 4 out of 5, signifying their significant impact on the implementation process. The diversity of legal systems and traditions across the African continent complicates the Harmonization of data protection laws, as each country must first reconcile its national laws with the Convention's provisions. Moreover, striking the right balance between national security and data privacy is delicate. Governments must ensure they have the tools to safeguard citizens and maintain public order while respecting their obligations under the Convention.

The diversity of legal systems and traditions across the African continent makes harmonizing data protection laws more complex. Each country must first reconcile its national laws with the Convention's provisions. The process can be lengthy and resource-intensive, as countries must navigate potential conflicts between the Convention and their existing legal frameworks. Given the importance of Harmonization for successfully implementing the Convention, this challenge has been assigned a magnitude of 4 out of 5.

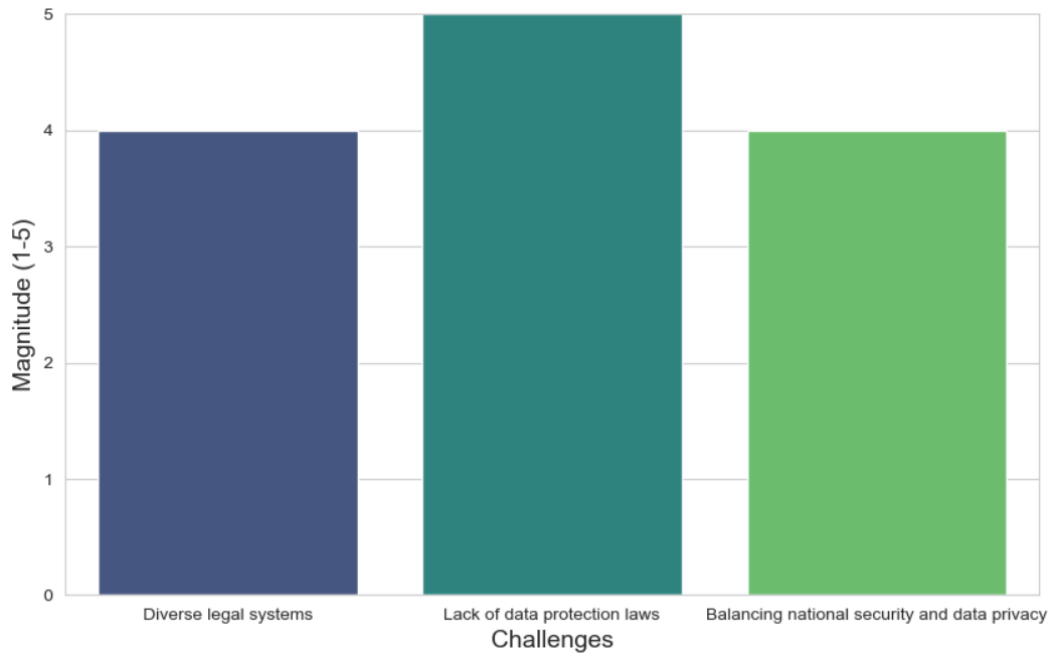

*Figure 1 Legal and Regulatory Challenges Challenges.*

Moreover, Many African countries are still in the early stages of developing their data protection legal frameworks. The absence of comprehensive data protection laws can hinder the Convention's effective implementation, as countries without such laws may lack the necessary legal tools to enforce its provisions. This challenge also creates uncertainty for businesses and organizations about their data privacy and protection responsibilities. The widespread lack of comprehensive data protection laws and its direct impact on the Convention's implementation has led to the assignment of the highest magnitude, 5 out of 5.

Striking the right balance between national security and data privacy is a delicate task for governments. Ensuring the necessary tools to safeguard citizens and maintain public order while respecting obligations under the Convention is challenging. This balance is essential in cross-border data transfers and cooperation between law enforcement and intelligence agencies. Due to the complexities in finding this balance and its implications for individual privacy rights and national security, this challenge has been assigned a magnitude of 4 out of 5.

### 3.5.2. Comparing Technical And Infrastructural Challenges

The bar chart in Figure 2 visually represents the technical and infrastructural challenges faced in implementing the AUDPC.

The chart shows that the "Cybersecurity threats" challenge has the highest magnitude (5 out of 5), indicating that it is a critical issue that needs to be addressed. This rating is justified by the increasing prevalence of cyber attacks targeting African countries and the potential impact of data breaches on the effectiveness of data protection laws and public trust in digital services.

The "Digital divide" and "Emerging technologies" challenges have a magnitude of 4 out of 5, signifying their significant impact on the implementation process. The magnitude of the "Digital divide" challenge is justified by the uneven access to ICT across the continent, which can hinder the effective implementation of the Data Privacy Convention and exacerbate existing social, economic, and political inequalities. Addressing this challenge requires substantial investments in expanding ICT infrastructure and digital literacy programs.

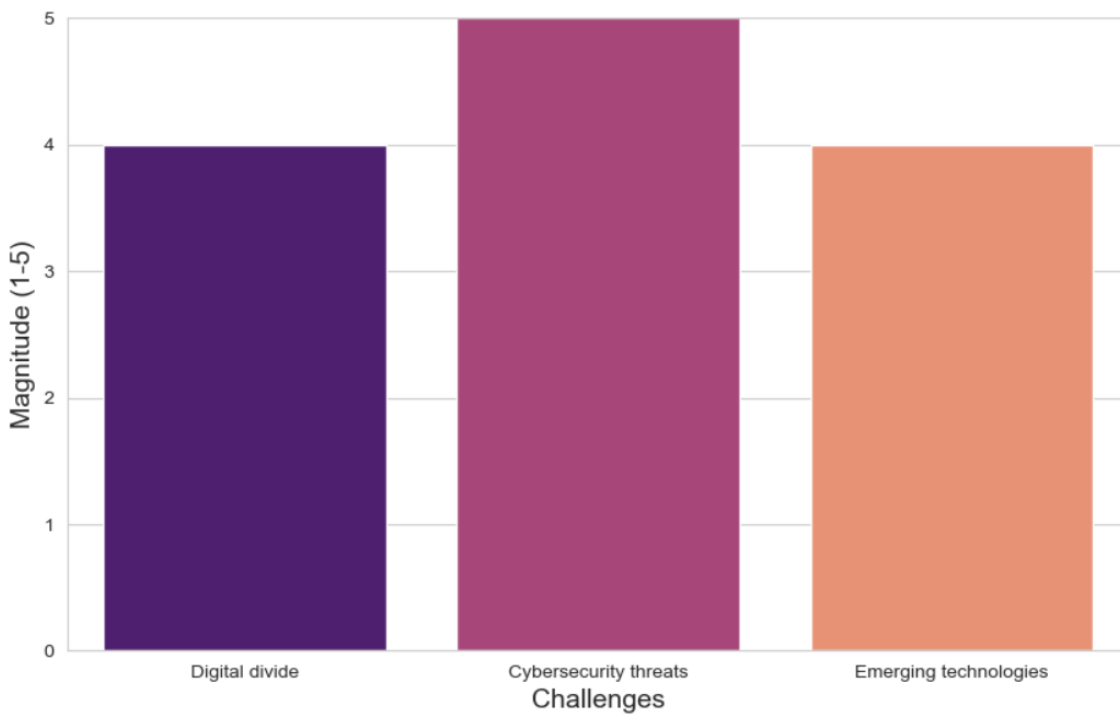

Figure 2 Technical and Infrastructural Challenges.

The "Emerging technologies" and the potential implications of technologies such as artificial intelligence, the Internet of Things, and big data analytics on data privacy justify the challenge's magnitude. These technologies often involve collecting, processing, and analyzing vast amounts of personal information, raising concerns about fairness, transparency, and accountability. To address this challenge, countries must ensure that their data protection laws and regulations are flexible enough to accommodate technological advancements while providing robust privacy protections.

By examining Figure 2 visualization and the rationale behind the assigned magnitudes, we gained deeper insight into the array of technical and infrastructural obstacles accompanying the Data Privacy Convention's implementation. Addressing these challenges is crucial to successfully achieving the Convention's objectives.

### 3.5.3. Capacity-Building And Awareness Challenges

The bar chart in Figure 3 visually represents the capacity-building and awareness challenges faced in implementing the AUDPC.

From the chart, we can observe that the "Limited resources for DPAs" and "Capacity building" challenges have a magnitude of 4 out of 5, indicating that they are significant issues that need to be addressed. The magnitude of the "Limited resources for DPAs" challenge is justified by the crucial role that data protection authorities (DPAs) play in enforcing data privacy laws and ensuring compliance with the Data Privacy Convention. Many African countries face challenges in providing adequate resources to establish and maintain effective DPAs, which can hinder their ability to carry out their responsibilities effectively.

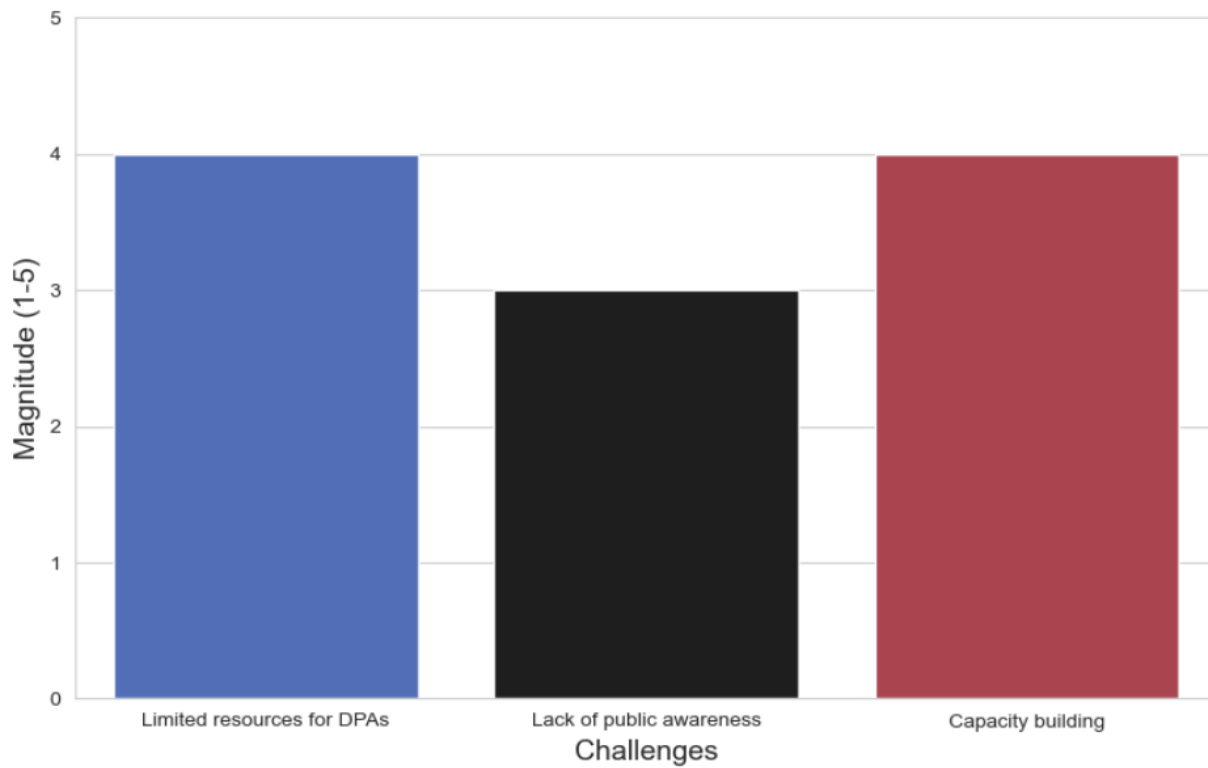

Figure 3 Capacity Building and Awareness Challenges.

The "Capacity building" challenge's magnitude (4 out 5) is justified by the need to ensure that governments, businesses, and civil society organizations can effectively implement the Data Privacy Convention. This includes training legal professionals, law enforcement officers, IT specialists, and other relevant stakeholders on data protection principles, best practices, and the requirements of the Convention. Adequate capacity building is essential for fostering a culture of privacy within organizations and ensuring stakeholders are equipped to address the evolving challenges of data privacy and protection.

The "Lack of public awareness" challenge has a magnitude of 3 out of 5, signifying its impact on the implementation process. This rating is justified by the importance of raising public awareness about data privacy rights to ensure the successful implementation of the Data Privacy Convention. A lack of understanding can lead to under-reporting of privacy violations and reduced demand for more robust data protection measures. Efforts to increase public awareness may include developing educational materials and campaigns, hosting workshops, and leveraging social media and other digital platforms to reach a broad audience.

Examining Figure *3* visualization and the rationale behind the assigned magnitudes enables a deeper comprehension of the diverse capacity-building and awareness obstacles related to implementing the Data Privacy Convention. This highlights the significance of tackling each challenge to achieve the Convention's objectives successfully.

### 3.5.4. Harmonization And Cross-Border Cooperation Challenges

The bar chart in Figure 4 visually represents the harmonization and cross-border cooperation challenges faced in implementing the AUDPC.

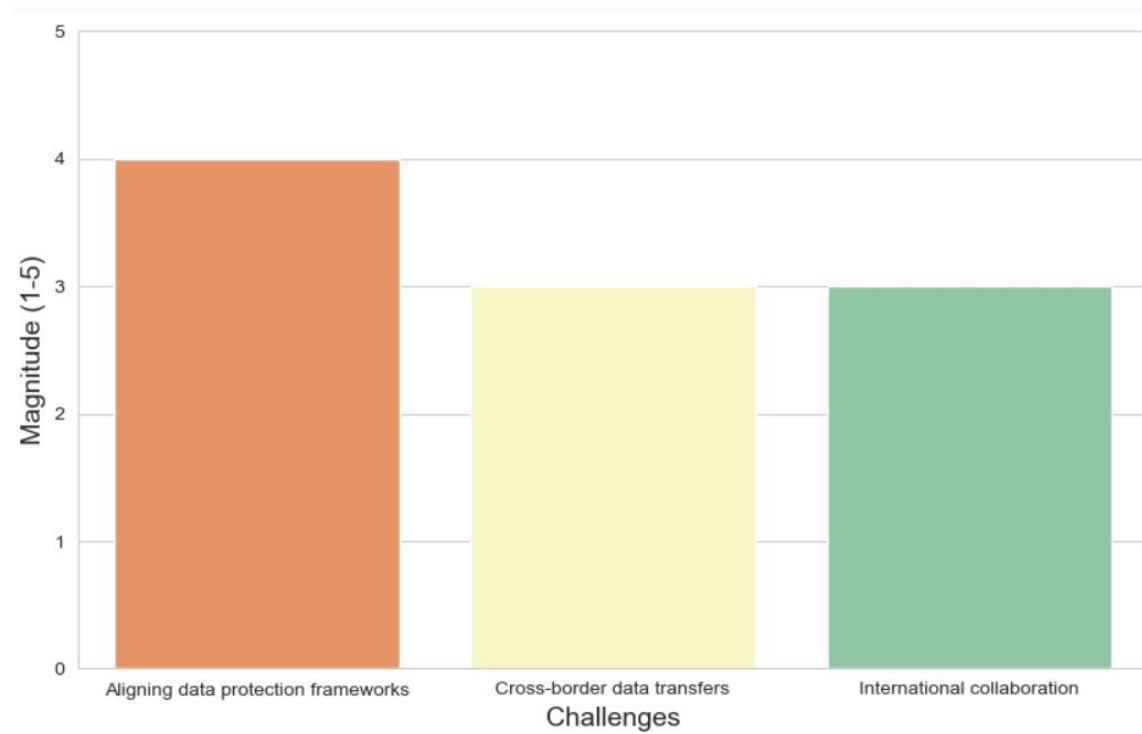

Figure 4 Harmonization and Cross-border Cooperation Challenges.

The chart shows that the "Aligning data protection frameworks" challenge has the highest magnitude (4 out of 5), indicating that it is a significant issue that needs to be addressed. The magnitude of this challenge is justified by the need for consistency in privacy protection standards and the facilitation of cross-border data flows within the continent. Achieving this Harmonization requires reconciling differences in legal traditions, addressing existing data protection laws gaps, and ensuring compatibility with national legislation.

The "Cross-border data transfers" and "International collaboration" challenges have a magnitude of 3 out of 5, signifying their impact on the implementation process. The magnitude of the "Cross-border data transfers" challenge is justified by the growing need for mechanisms to facilitate cross-border data transfers while ensuring adequate privacy protection. This may involve implementing measures such as adequacy decisions, standard contractual clauses, and binding corporate rules.

The "International collaboration" challenge's magnitude is justified by the importance of collaborating with international partners and global organizations to ensure that African countries can effectively participate in the global digital economy and protect the privacy rights of their citizens. This collaboration may involve sharing best practices, providing technical assistance, and engaging in policy dialogues to address common challenges and emerging data privacy and protection issues.

Through examining Figure *4* visualization and the rationale behind the assigned magnitudes, we gain a deeper comprehension of the diverse Harmonization and cross-border cooperation challenges linked to implementing the Data Privacy Convention. By emphasizing the significance of addressing each problem, we underscore the necessity of achieving the Convention's objectives successfully.

## 4. AUDPC Future Directions

To effectively address the challenges associated with implementing the AUDPC and protect personal information across the continent, it is essential to focus on several critical future directions. This section will explore the importance of strengthening legal and regulatory frameworks, enhancing technical and infrastructural capacities, fostering capacity-building and awareness initiatives, promoting Harmonization and cross-border cooperation, and engaging with global data protection trends and developments.

### 4.1. Strengthening Legal And Regulatory Frameworks

Strengthening legal and regulatory frameworks in African countries is crucial for creating a comprehensive data protection environment. This entails reviewing and updating existing data privacy laws, closing gaps in current legislation, and adopting new regulations in line with the Data Privacy Convention's principles. Governments should engage with stakeholders, including the private sector, civil society, and academia, to ensure that the updated frameworks address the needs and concerns of all parties and promote a culture of privacy.

### 4.2. Enhancing Technical And Infrastructural Capacities

Enhancing technical and infrastructural capacities is vital for protecting personal information in the digital age. African countries should invest in improving their ICT infrastructure and fostering innovation in privacy-enhancing technologies. This may involve supporting research and development initiatives, creating partnerships with the private sector, and promoting open-source technologies to ensure accessibility and affordability.

### 4.3. Fostering Capacity-Building And Awareness Initiatives

To address capacity-building and awareness challenges, African countries should invest in initiatives that build the skills and knowledge of stakeholders, including government officials, law enforcement officers, and legal professionals. This can be achieved through workshops, training programs, and collaborative learning opportunities. In addition, public awareness campaigns should be developed and implemented to educate citizens about their data privacy rights and to promote a culture of privacy.

### 4.4. Promoting Harmonization And Cross-Border Cooperation

Promoting Harmonization and cross-border cooperation is crucial for facilitating the free flow of data and ensuring consistent privacy protection standards across the continent. African countries should engage in dialogue and collaboration to align their data protection frameworks and establish mechanisms for cross-border data transfers. This may involve creating regional working groups, sharing resources and expertise, and participating in international forums and initiatives related to data privacy.

### 4.5. Engaging With Global Data Protection Trends And Developments

African countries should actively engage with global data protection trends and developments to ensure their data privacy laws and regulations remain updated and effective. This may involve participating in international conferences, collaborating with global organizations, and monitoring emerging technologies and their implications for data privacy. By staying informed of global trends and developments, African countries can better anticipate and address the challenges of protecting personal information in an increasingly interconnected world.

Concentrating on these prospective paths, African nations can establish a strong, all-encompassing data protection landscape that secures their citizens' privacy rights, builds confidence in digital services, and encourages economic development and innovation.

## 5. Conclusion

Implementing the AUDPC presents challenges and opportunities for the continent. This paper has examined the legal, regulatory, technical, infrastructural, capacity building, and awareness challenges and Harmonization and cross-border cooperation challenges African countries face in this context. To address these challenges and ensure the successful implementation of the Convention, it is essential to focus on crucial future directions, such as strengthening legal and regulatory frameworks, enhancing technical and infrastructural capacities, fostering capacity building and awareness initiatives, promoting Harmonization and cross-border cooperation, and engaging with global data protection trends and developments.

The key findings of this paper include the identification of diverse legal systems and traditions, the lack of comprehensive data protection laws, and the need to balance national security and data privacy as significant legal and regulatory challenges. Technical and infrastructural challenges encompass the digital divide, cybersecurity threats, and the implications of emerging technologies on data privacy. Capacity building and awareness challenges include limited resources for data protection authorities, lack of public awareness about data privacy rights, and the need for capacity building in data privacy and protection. Lastly, harmonization and cross-border cooperation challenges involve aligning data protection frameworks across the continent, establishing mechanisms for cross-border data transfers, and collaborating with international partners and global organizations.

Based on the analysis and findings presented in this paper, the following suggestions are proposed for the African Union and its member states to address the challenges and facilitate the successful implementation of the Data Privacy Convention:

- Strengthen legal and regulatory frameworks by reviewing and updating existing data privacy laws, closing gaps in current legislation, and adopting new regulations in line with the Data Privacy Convention's principles.
- Enhance technical and infrastructural capacities by investing in ICT infrastructure, fostering innovation in privacy-enhancing technologies, and supporting research and development initiatives.
- Foster capacity building and awareness initiatives by developing and implementing training programs, workshops, and collaborative learning opportunities for stakeholders and launching public awareness campaigns to educate citizens about their data privacy rights.
- Promote Harmonization and cross-border cooperation by engaging in dialogue and collaboration to align data protection frameworks, establish mechanisms for cross-border data transfers, and participate in international forums and initiatives related to data privacy.
- Engage with global data protection trends and developments by participating in international conferences, collaborating with global organizations, and monitoring emerging technologies and their implications for data privacy.

Through the implementation of these recommendations, the African Union and its member states can effectively address the challenges associated with the Data Privacy Convention, safeguard the privacy rights of their citizens, and foster trust in digital services, ultimately promoting economic growth and innovation across the continent.

[20]    D. Coleman, Digital colonialism: The 21st century scramble for Africa through the extraction and control of user data and the limitations of data protection laws, Mich. J. Race \& L. 24 (2018) 417.

[21]    Deloitte, Privacy is Paramount: Personal Data Protection in Africa, Deloitte. (2017) 1–12. https://www2.deloitte.com/content/dam/Deloitte/za/Documents/risk/za_Privacy_is_Paramount-Personal_Data_Protection_in_Africa.pdf.
Page **15** of **15**